\documentclass[prb,aps,twocolumn,draft,tightenlines,floatfix,superscriptaddress,showpacs]{revtex4}
\usepackage{psfig}
\usepackage{amssymb}
\begin{document}
\title{Study of the formation and decay of electron-hole plasma
  clusters in a direct-gap semiconductor CuCl}
\author{L. Jiang}
\affiliation{Structure Research Laboratory, University of Science \&
  Technology of China, Academia Sinica,  Hefei, Anhui, 230026, China}
\affiliation{Department of Physics, University of Science \&
  Technology of China, Hefei, Anhui, 230026, China}
\altaffiliation{Mailing Address.}
\author{M. W. Wu}
\thanks{Author to whom correspondence should be addressed}
\email{mwwu@ustc.edu.cn}
\affiliation{Structure Research Laboratory, University of Science \&
  Technology of China, Academia Sinica,  Hefei, Anhui, 230026, China}
\affiliation{Department of Physics, University of Science \&
  Technology of China, Hefei, Anhui, 230026, China}
\author{M. Nagai}
\affiliation{Department of Physics, Kyoto University, Kyoto 606-8502, Japan}
\author{M. Kuwata-Gonokami}
\affiliation{Department of Applied Physics, University of Tokyo, Hongo, Bunkyo-ku, Tokyo 113-8656, Japan}
\date{\today}
\begin{abstract}

The master equation for the cluster-size distribution function is
solved numerically to investigate the electron-hole droplet formation
claimed to be discovered in the direct-gap CuCl excited by picosecond
laser pulses [Nagai {\em et al.}, Phys. Rev. Lett. {\bf 86}, 5795
(2001); J. Lumin. {\bf 100}, 233 (2002)]. Our result shows that for
the excitation in the experiment, the average number of pairs per
cluster (ANPC) is only around 5.2, much smaller than that (10$^6$
typically for Ge) of the well studied electron-hole droplet in
indirect-gap semiconductors such as Ge and Si.
\end{abstract}
\pacs{71.35.Ee, 71.35.Lk} 

\maketitle

 In photoexcited semiconductors, electrons and holes are generated by
the photon absorption and interact with each other via the Coulomb
interaction. When the exchange- and correlation-induced attractions
among the high density photogenerated carriers are strong enough to
overcome the fermionic repulsive pressure and to reduce the chemical
potential below the exciton energy, electron-hole droplets (EHDs)
appear as first order transition.\cite{lvk} Investigation of the
formation kinetics of EHD provides an opportunity to understand the
quantum many-body effects.  In semiconductor with an indirect gap and
correspondingly long optical lifetimes, such as Ge and Si, EHDs have
been investigated both theoretically and experimentally in great
detail.\cite{sige} Within the coexistence regime, relatively large EHD
clusters are formed and the kinetics of the EHD formation is well
understood in terms of the master equation\cite{sil,haug} where the
ANPC $\langle n\rangle$ is very large (for Ge $\langle n\rangle$ is
about 10$^6$).

\begin{figure}[htb]
  \psfig{file=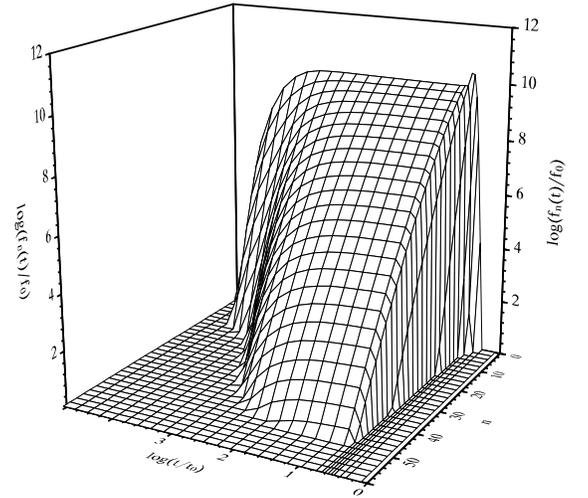,width=10.cm,height=9cm,angle=0}
  \caption{Cluster concentration versus time and number of e-h pairs
    per cluster for CuCl at $T=4.2$~K under 
    Gaussian pulse excitation, $t_r$=0.2~ps, $G_0=6.1\times
    10^{33}$~cm$^{-3}$s$^{-1}$,  and
    $\rho_0=2.0\times10^{20}$~cm$^{-3}$. $f_0=10^{10}$\ cm$^{-3}$.}
  \label{fig1}
\end{figure}

Unlike indirect gap semiconductors, in direct gap semiconductors the
situation is very complicated because the system cannot reach
quasiequilibrium during the carrier lifetime (typically in the order
of 1\ ns). Despite extensive experimental efforts, no decisive
evidence of the EHD formation has yet been reported for a very long
time.\cite{haug2} EHD formation has been predicted theoretically in
polar CdS and CdSe.\cite{rice1} Also, EHD-like behavior has been
observed in them experimentally in which time-resolved emission
measurements are performed.\cite{vbtim1,ssion1} This method has also
been applied to GaAs.  It reveals that electron-hole plasma created by
band-to-band excitation has extremely high temperature due to the
overheating caused by the reduction of band gap.  Very recently, Nagai
{\em et al.} used CuCl which has large binding energy and band gap
(binding energy of 1s-Z$_3$ exciton E$_{ex}$=213~meV and band-gap
energy E$_g$=3.396~eV) to overcome the problem. They therefore are
able to study the formation of EHD in CuCl via time-resolved emission
and pump-probe reflection measurements by femtosecond excitation in
the mid-infrared region under resonant exciton excitation.  They
reported that clear EHD-like behavior in CuCl has been
found.\cite{ku1,ku2}

However, due to the fact of the short lifetime of the carrier, one
expects only small clusters can be formed and sustained. Two decades
ago, Haug and Abraham performed a theoretical investigation of the EHD
formation in GaAs.\cite{haug1} They reported for the typical
picosecond excitation the ANPC is at most around 2 where there should
be no EHD formation can be found.  In this paper we use the master
equation theory developed by Haug and Abraham\cite{haug1} to
investigate the femtosecond excitation in CuCl in the experiments of
Nagai {\em et al.}.\cite{ku1,ku2} This theory has been proven to be 
very sucessful in dealing with the EHD formation in indirect gap semiconductors.\cite{sil,haug}
In this study, we focus on how big the
ANPC can be in this material and how this number is affected by
different parameters. 

\begin{figure}[htbp]
\psfig{file=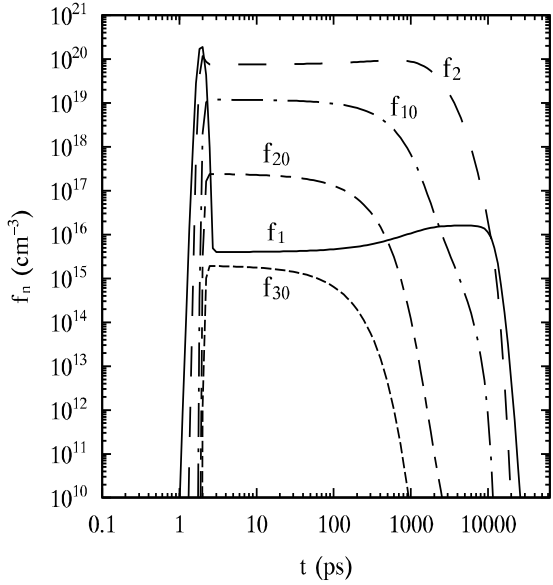,width=10.5cm,height=9.5cm,angle=0}
  \caption{Concentrations of selected clusters versus time for CuCl at
    T=4.2~K under Gaussian pulse excitation 
   as in figure 1.}
  \label{fig2}
\end{figure}

The concentration $f_n(t)$ to find at a given time t a cluster with $n$
e-h pairs per unit volume is given by the following master equation
equations:\cite{haug1} For $n>1$,
\begin{eqnarray}
&&\frac{\partial f_n}{\partial t}=j_{n-1}-j_n\\ &&j_n=g_n f_n-l_{n+1}
f_{n+1}
\end{eqnarray}
where $j_n$ is the net probability current between the clusters with n
and n+1 e-h pairs, $g_n$ is the gain rate of a cluster with n e-h
pairs and is approximated by
\begin{eqnarray}
g_n&=&4\pi R_n^2 v_0f_1 \\ v_0&=&\sqrt{k T/2 \pi m^{\ast}} \\
n&=&\frac{4\pi}{3}R_n ^3 \rho_0
\end{eqnarray}
$R_n$ is the radius of the cluster, $v_0$ is the thermal exciton
velocity, $\rho_0$ is the liquid density. The loss rate $l_n$ is the
sum of the evaporation rate $e_n$ and the recombination rate
$n/\tau_n$:
\begin{equation}
  l_n=e_n+n/\tau_n\ .
\end{equation}
 $e_n$ is given by the Richardson-Dushman current\cite{sil}
\begin{equation}
e_n=\gamma\left(\frac{m^\star kT}{2\pi
\hbar^2}\right)^{3/2}\frac{g_{n-1}}{f_1}\exp[(-\phi+c \sigma
n^{-1/3})\beta]
\end{equation}
$\gamma$ is the degeneracy of the exciton level, $\phi$ is the binding
energy of the liquid with respect to the exciton. $c\sigma n^{-1/3}$
is the correction of the binding energy with $\sigma$ denoting the surface energy of the liquid. 
Typically we select $\gamma$
=1. $\sigma=C\varphi \rho_0 r_0$ with $C\sim 1$, $\varphi$ standing for the condensate energy 
per electron-hole pair and
$r_0=(3/4\pi\rho_0)^{1/3}$.\cite{lvk}  The change of exciton (corresponding to $n=1$) is treated
separately:
\begin{equation}
\frac{\partial f_1}{\partial t}=G(t)-\sum_{n=1}^{\infty}
\frac{f_n}{\tau_n}-2j_1-\sum_{n=2}^{\infty}j_n
\end{equation}
where G(t) is the generation rate of excitons.

\begin{figure}[htb]
  \psfig{file=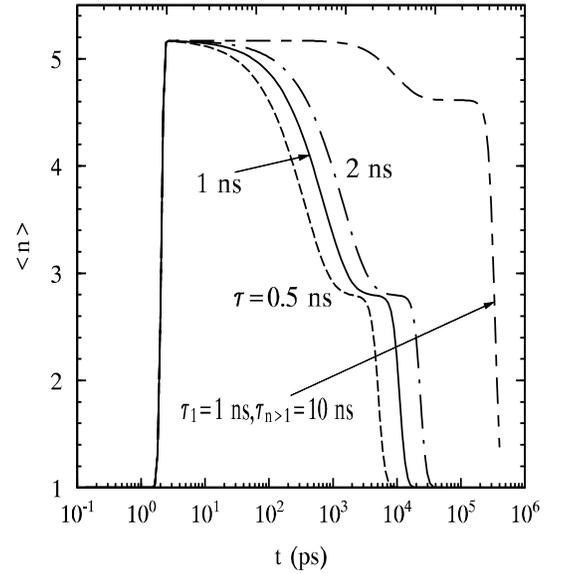,width=10.5cm,height=9.5cm,angle=0}
  \caption{ANPC versus time for
    CuCl at T=4.2~K under Gaussian pulse excitation.  $t_r$=0.2~ps, $G_0=6.1\times 10^{33}$~cm$^{-3}$s$^{-1}$, $\rho_0=2.0\times10^{20}$~cm$^{-3}$. Solid curve: $\tau=1$\ ns; Dashed curve: $\tau= 0.5$\ ns; Dot-dashed curve: $\tau $=2~ns.}
  \label{fig3}
\end{figure}

  We solve above equations numerically for the example of CuCl. The
parameters for CuCl at 4.2~K are $m^{\ast}$=2.0$m_0$ and $\phi=5.12
\times 10^{-14}$~erg.\cite{d1} The value of the binding energy $\phi$ is taken
to be the biexciton binding energy which is the lower limit of the binding energy. Our computation indicates that the 
result is almost unchanged if one increases $\phi$ by one order of magnitude.
  We obtain $\rho_0$ from
$r_s$=$\sqrt[3]{3/4\pi na_0^3}$, where $a_0$ is the Bohr radius and
$r_s$=1.8. \cite{ku1} $\sigma=0.54$~erg$\cdot$cm$^{-2}$. In our calculation we assume $\tau_n\equiv
\tau$.  According to the experiment, we choose a Gaussian pulse:
\begin{equation}
  G(t)=G_0 e^{-(t-t_0)^2/t_r^2}
\end{equation}
where the pulse width $t_r=0.2$\ ps, and the peak generation rate is
$G_0=6.1\times10^{33}$~cm$^{-3}$s$^{-1}$, corresponding to a
excitation density of 4.3 mJ$\cdot$cm$^{-2}$.  The main results of our
calculation are plotted in Figs. 1-3.

Figure 1 shows the development of the distribution of cluster
distribution. For the short pulse, the exciton concentration $f_1(t)$ rapidly reaches its 
maximum near the pulse center, $t_0=2$\ ps. Then it decreases quickly due to the formation of the
clusters. It is also noted from the figure that within 1\ ps after the center of the pulse, 
clusters of all sizes reaches to their maximum values. As any cluster with size $n\ge 2$ comes from 
the exciton, $f_1(t)$ sharply drops 5 orders of magnitude from its peak within
that period of time.  After that, because of the recombination and
the evaporation, almost all clusters start to decay, except for some small clusters (say $n=1$ and 2) which
increase a little bit due to the evaporation from bigger clusters.
 The lifetime of large cluster (say $n=50$) is much shorter
than that of the small one as the later gets compensation from the decay of
the larger clusters and also the recombination rate $n/\tau_n$ is propotional to $n$.  
After 10~ns almost all
cluster concentrations are below  $10^{10}$~cm$^{-3}$. This
can be seen more quantitatively in Fig. 2 where the
concentration of some selected clusters are plotted as functions of time.

In order to see in average how large the cluster can be under the experiments by Nagai {\em et al.}, we 
plot in Fig.\ 3 the ANPC $\langle n\rangle$ as function of time for three different
lifetimes $\tau$. $\langle n\rangle$ can be calculated from
\begin{equation}
  \langle n\rangle ={\sum_{n=1}^{\infty} n f_n\over \sum_{n=1}^{\infty}
    f_n}\ .
\end{equation}
One notices from the figure that for the excitation in the experiment, ANPC is only 5.2, orders of
magnitude smaller than the number in the EHD of indirect semiconductor such as Ge and Si. This result is consistent 
with the regular expectation of the excitation in the direct gap semiconductors. 
Moreover, if one assumes that the larger clusters are stable and therefore $\tau_n$ ($n\ge2$) is larger than
$\tau_1$. We demonstrate in the same figure the result when $\tau_1=1$\ ns but $\tau_n=10$\ ns when $n\ge2$.
One finds that this does not affect the ANPC, although the metastable time is much longer. We further find that if we
increase the excitation $G_0$ by one order of magnitude, we can only double the ANPC. These results
indicate that the feature of low ANPC is very robust.
These results indicate that what measured in CuCl by Nagai {\em et al.}\cite{ku1,ku2} 
may not come from the EHD formed from exciton gas, but instead possibly  come
from some bubbles of excitons in metallic  liquid.\cite{ku2}

MWW is supported by the ``100 Person Project'' of
Chinese Academy of Sciences and Natural Science Foundation
of China under Grant No. 10247002. He would also like to thank S. T. Chui at Bartol Research Institute,
University of Delaware for hospitality. LJ would like to thank M. Q. Weng for his help during this work.
MWW, MN and MK is supported in part by the Solution Oriented Research for
Science and Technology  (SORST)  of the Japan Science and Technology
Corporation (JST).

\end {document}